\documentclass{article}  
\usepackage{bigsky2009}
\usepackage{graphicx}

\newcommand{\lsim}{\,{\buildrel < \over {_\sim}}\,}

\frompage{000} \topage{000}                                              %|
\def\rightmark{Jet fragmentation in STAR}
\def\leftmark{E. Bruna}
 
\title{Jet fragmentation in STAR going from p+p to Au+Au} 
\authors{
{Elena Bruna$^1$, for the STAR Collaboration 
}\\[2.812mm]
{\normalsize
\hspace*{-8pt}$^1$ Physics Department, Yale University \\ 
272 Whitney Avenue, New Haven, CT-06511 USA\\[0.2ex] 
}}
 
\abstract{
Jet fragmentation functions provide insight into jet structure and are expected to be modified by the nuclear medium in A+A collisions with respect to p+p reference measurements. If jet reconstruction is unbiased then a softening of the fragmentation functions is expected and should be observed in Au+Au collisions at RHIC.
In these proceedings we present measurements of fragmentation functions in p+p for charged particles for different jet finding algorithms; these measurements are understood and therefore can be used as a reference for comparison with Au+Au results.
We report the effect of background and its fluctuations on jet reconstruction in Au+Au collisions, estimated by using the jet algorithms on simulated Pythia jets embedded in real Au+Au events. Finally, measurements of fragmentation functions for jets reconstructed in Au+Au events and their comparison to the p+p baseline are presented and discussed.
 }


\begin{document}
 
\maketitle
\setcounter{page}{1}
\section{Introduction}
High-$p_T$ physics via di-hadron correlations has been a successful tool to probe the dense medium created in heavy-ion collisions at RHIC. Nevertheless, well-known geometrical biases~\cite{renk} provide limitations to such measurements. 
Full reconstruction of jets in heavy-ion collisions should allow direct access to the parton kinematics independent of the fragmentation process in an unbiased way.
%Such jet analyses are possible in STAR with the Time Projection Chamber and the Electromagnetic Calorimeter.
The first $p_T$ spectrum of jets reconstructed with a cone algorithm in p+p collisions at RHIC energies was measured by STAR with the 2003-2004 data, showing good agreement with NLO pQCD calculations over the measured $p_T$ range up to 50 GeV~\cite{starjets1,starjets2}.
Recent results from STAR~\cite{joern,sevil} show the feasibility  of full jet reconstruction at RHIC in the high-multiplicity environment of Au+Au collisions. 
The underlying background and its large fluctuations on an event-by-event basis make these measurements a challenge in the high-multiplicity environment at RHIC.
Jet fragmentation functions provide insight into the structure of jets and are expected to be modified by the nuclear medium with respect to p+p reference measurements~\cite{wiedemann}. 
An expected softening of the fragmentation functions, i.e. a distortion of the so called ``hump-backed'' plateau~\cite{borghini}, should be observable in Au+Au with an unbiased jet population.

%In these proceedings we will present measurements of fragmentation functions in p+p for charged particles for different jet finding algorithms; these measurements are under control and therefore pave the way for a comparison with Au+Au results.
%We will report the effect of background and its fluctuations in Au+Au collisions, estimated by running the jet algorithms on simulated Pythia jets embedded in real Au+Au events. Finally, measurements of fragmentation functions for jets reconstructed in Au+Au events and their comparison to the baseline p+p are presented and discussed.

\section{Technical aspects in p+p and Au+Au analyses}
\subsection{Event selection}
The STAR detectors employed for jet analyses are the Time Projection Chamber (TPC) that reconstructs charged particles and the Barrel Electromagnetic Calorimeter (BEMC) for the neutral energy.
Corrections for double-counting of electrons and hadronic energy deposition in the BEMC are applied. The following trigger setups were used:\\
%\begin{itemize}
%\item 
(i) Jet Patch Trigger (JP) in p+p: year 6 run, with the transverse energy in a BEMC cluster $\Delta \eta \times \Delta \phi=1 \times 1$ $E_T>$8 GeV (in addition to p+p minimum bias condition). The total luminosity is  8.7 pb$^{-1}$.
\\
(ii) High Tower Trigger (HT) in p+p: year 6 run, transverse energy in a single BEMC tower $\Delta \eta \times \Delta \phi=0.05 \times 0.05$ $E_T>$5.4 GeV (in addition to p+p minimum bias condition). The total luminosity is  11 pb$^{-1}$.
%(iii) High Tower Trigger (HT) in p+p: year 6 run, transverse energy in a single BEMC tower $\Delta \eta \times \Delta \phi=0.05 \times 0.05$ $E_T>$5.4 GeV (in addition to p+p minimum bias condition). The total luminosity is  11 pb$^{-1}$.
\\
(iii) Minimum Bias Trigger (MB) in Au+Au: year 7 run, with coincidence condition between the two ZDCs.
\\
(iv) High Tower Trigger (HT) in Au+Au: year 7 run, transverse energy in a BEMC cluster $\Delta \eta \times \Delta \phi=0.1 \times 0.1$ $E_T>$7.8 GeV (in addition to MB condition). The total luminosity is  500 $\mu$b$^{-1}$, corresponding to a p+p equivalent of 19.6 pb$^{-1}$.
\\

%To avoid double-counting of electrons which leave signals both in the TPC and in the BEMC, the BEMC energy that matches electron candidates (identified with p/E criteria) is rejected and only the corresponding track $p_T$ is kept. In addition, since a charged hadron is expected to deposit part of its energy in the calorimeter, the energy of one MIP is subtracted from a tower hit by a charged hadron.

\subsection{Jet-finding algorithms}
Jet-Finding algorithms cluster charged tracks and neutral towers into jets, which are experimentally defined as sprays of collimated particles in the $\eta-\phi$ plane.    
Cone and recombination algorithms (see Ref.~\cite{reviewjets} for more details) are the two main jet-finding tools. 
Cone algorithms start from those bins (``seeds'') in the $\eta-\phi$ plane whose corresponding energy (from tracks and towers) is above a given threshold. 
For each seed, the algorithm searches for other particles in a cone of radius R ($R=\sqrt{(\Delta \phi^2 + \Delta \eta^2)}$). 
Cone jet-finders can be optimized by introducing iteration procedures and splitting/merging techniques~\cite{reviewjets}. The infrared safety is an issue in jet algorithms and is partially solved by adding extra midpoint seeds. A proper solution is to use a seedless cone algorithm, like SISCone (Seedless Infrared Safe Cone)~\cite{SISc}, which is part of the FastJet~\cite{fastjet} package.
Recombination algorithms are seedless and cluster particles according to their spacial separation with a distance scale R and weight criteria.
They are not generally bound to a circular structure. In the ``$k_T$'' algorithm the weight is given by the particle $p_T$ and therefore soft particles will tend to cluster among themselves first.
In contrast, in the ``anti-$k_T$'' algorithm~\cite{antikt} the distance is weighted by the inverse $p_T$ and hard particles are clustered together long before soft particles. Recombination algorithms are collinear and infrared safe.

\section{Fragmentation functions in p+p}
\begin{figure}
\vspace{-3mm}
\centering
\resizebox{0.49\textwidth}{!}{  \includegraphics{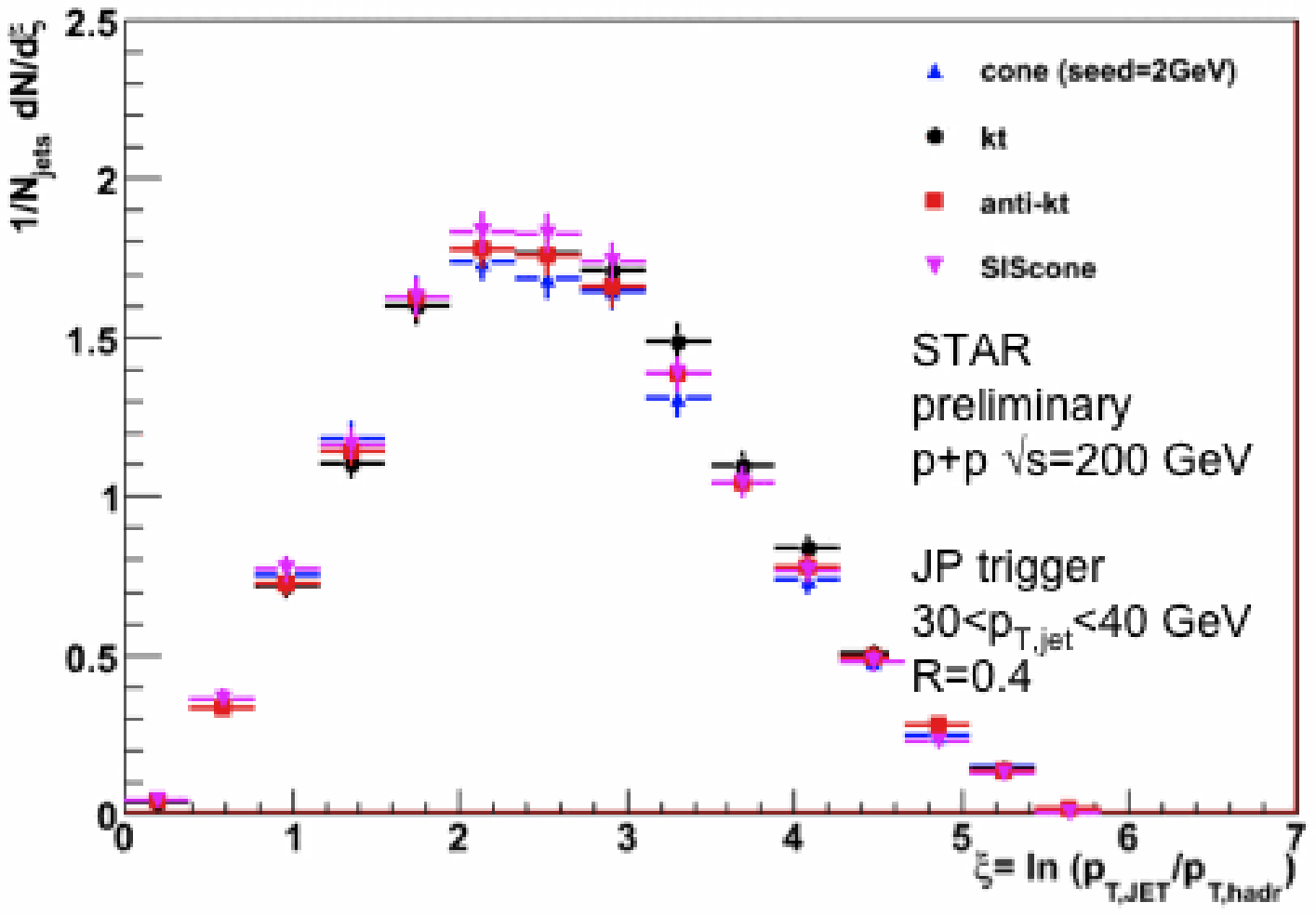}}
\resizebox{0.49\textwidth}{!}{  \includegraphics{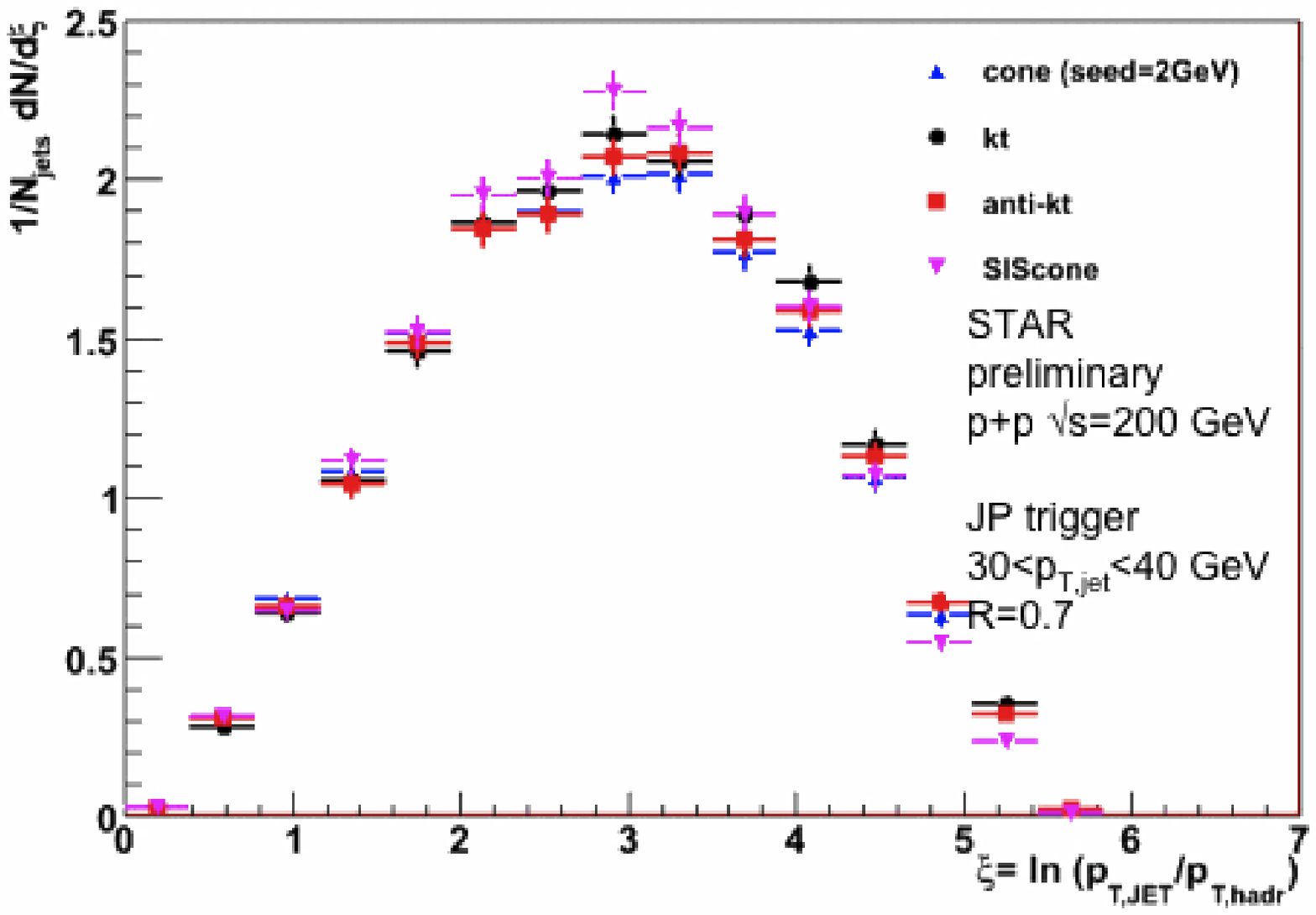}}

\vspace{-3mm}
\caption{Uncorrected $\xi$ distributions for four different jet algorithms for $30<p_{T,jet}<40$ GeV with JP data, R=0.4 (left) and R=0.7 (right).}
\label{fig:FF}
\end{figure}

The variable we use to describe the fragmentation function is $\xi=\ln(p_{T,jet}/p_{T,h})$, where $p_{T,jet}$ is the reconstructed transverse momentum of the jet and $p_{T,h}$ is the transverse momentum of the charged hadron in the jet. As an alternative to $z=p_{T,h}/p_{T,jet}$, $\xi$ is used to focus on the low-$p_T$ part of the fragmentation function.
We measured fragmentation functions from p+p JP data using four jet-finder algorithms (Seeded cone, ``$k_T$'', ``anti-$k_T$'', SISCone) for jet energies ranging from 10 GeV to 40 GeV and above and for two choices of the jet radius, R=0.4 and 0.7~\cite{mark,elena}.
It is known from Tevatron data that almost all the jet energy is contained in R=0.7 and more than $70\%$, increasing with the jet energy, lies within R$<$0.3.  R=0.4 was used for the purpose of comparing p+p fragmentation functions with Au+Au results, where a small jet radius is necessary to suppress the huge background.
Fig.~\ref{fig:FF} shows $\xi$ distributions for $30<p_{T,jet}<40$ GeV with R=0.4 (left) and R=0.7 (right).
Despite conceptual differences in the jet-finding algorithms studied, they agree over the range of jet energies explored.
The observation that jet algorithms show similar performance with different jet radii is an indication that second-order effects like out-of-cone radiation are minor at RHIC energies, otherwise jet algorithms could behave differently especially in case of small jet radii.

\section{Au+Au results}
\subsection{Background effect in jet-finding in Au+Au}
\begin{figure}
\vspace{-3mm}
\centering
\resizebox{0.9\textwidth}{!}{  \includegraphics{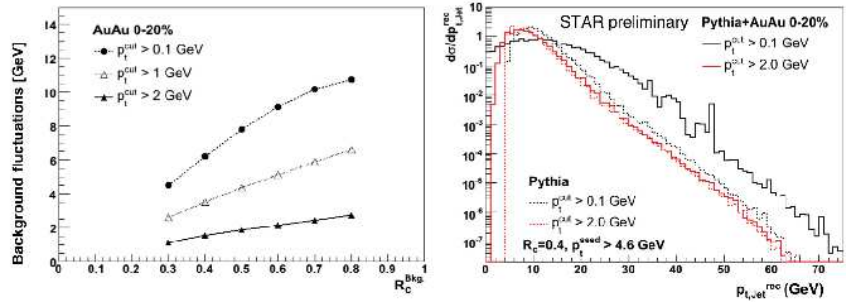}}
%\resizebox{0.9\textwidth}{!}{  \includegraphics{BkgFlu2.eps}}

\vspace{-3mm}
\caption{Left plot: width of background fluctuations as a function of jet radius for three choices of $p_t^{cut}$. Right plot: reconstructed Pythia jet spectrum with and without embedding in 0-20$\%$ MB Au+Au events at $\sqrt s_{NN}=200$ GeV.}
\label{fig:BkgFlu}
\end{figure}

As already mentioned, the background is the most crucial aspect in jet-finding in heavy-ion collisions.
Possible ways to suppress the background can be to reduce the jet radius and require a minimum transverse momentum ($p_t^{cut}$) for a track or a BEMC tower to be included in the jet.
On an event-by-event basis, the background contribution in the jet area is calculated as the mean $p_T$ in out-of-cone areas in the $\eta-\phi$ plane and is of the order of 45 GeV for R=0.4 and no $p_t^{cut}$. In addition to its magnitude, that is subtracted, the background exhibits significant fluctuations in $\eta-\phi$ regions on an event-by-event basis.
Background fluctuations in these data can be estimated using a Gaussian background model, with the width decreasing from $~6$ GeV with no  $p_t^{cut}$ to $~1.5$ GeV with $p_t^{cut}=2$ GeV in areas of R=0.4 (Fig.~\ref{fig:BkgFlu}, left).
Background fluctuations in Au+Au, on top of the steeply falling jet spectrum, result in a shift towards higher energies of the jet $p_T$ spectrum with respect to p+p. This is shown in  Fig.~\ref{fig:BkgFlu} (right), where spectra of Pythia jets reconstructed with a seeded cone algorithm (dashed lines) are compared to spectra obtained by running the same jet-finding algorithm on Pythia (version 8.1) jets embedded in real Au+Au events (solid lines). The effect of background fluctuations is dramatic with no $p_t^{cut}$ (black lines) and is substantially reduced by applying a $p_t^{cut}=2$ GeV (red lines).
This observation has important consequences in fragmentation function analyses, especially in view of comparing p+p and Au+Au measurements, where similar jet populations have to be selected in order to investigate possible modifications of the fragmentation functions. The most practical solution is to apply a  $p_t^{cut}=2$ GeV on both p+p and Au+Au events, that should result in a selection of jets with similar energies in p+p and Au+Au based on Pythia studies. The results of this analysis are reported in the next section. Alternatively, it is possible to avoid a $p_t^{cut}$ by unfolding the background fluctuations from the fragmentation functions and extract a distribution that is directly comparable to p+p. This work is in progress.

\subsection{Fragmentation functions in Au+Au}
\begin{figure}
\vspace{-3mm}
\centering

\resizebox{0.44\textwidth}{!}{  \includegraphics{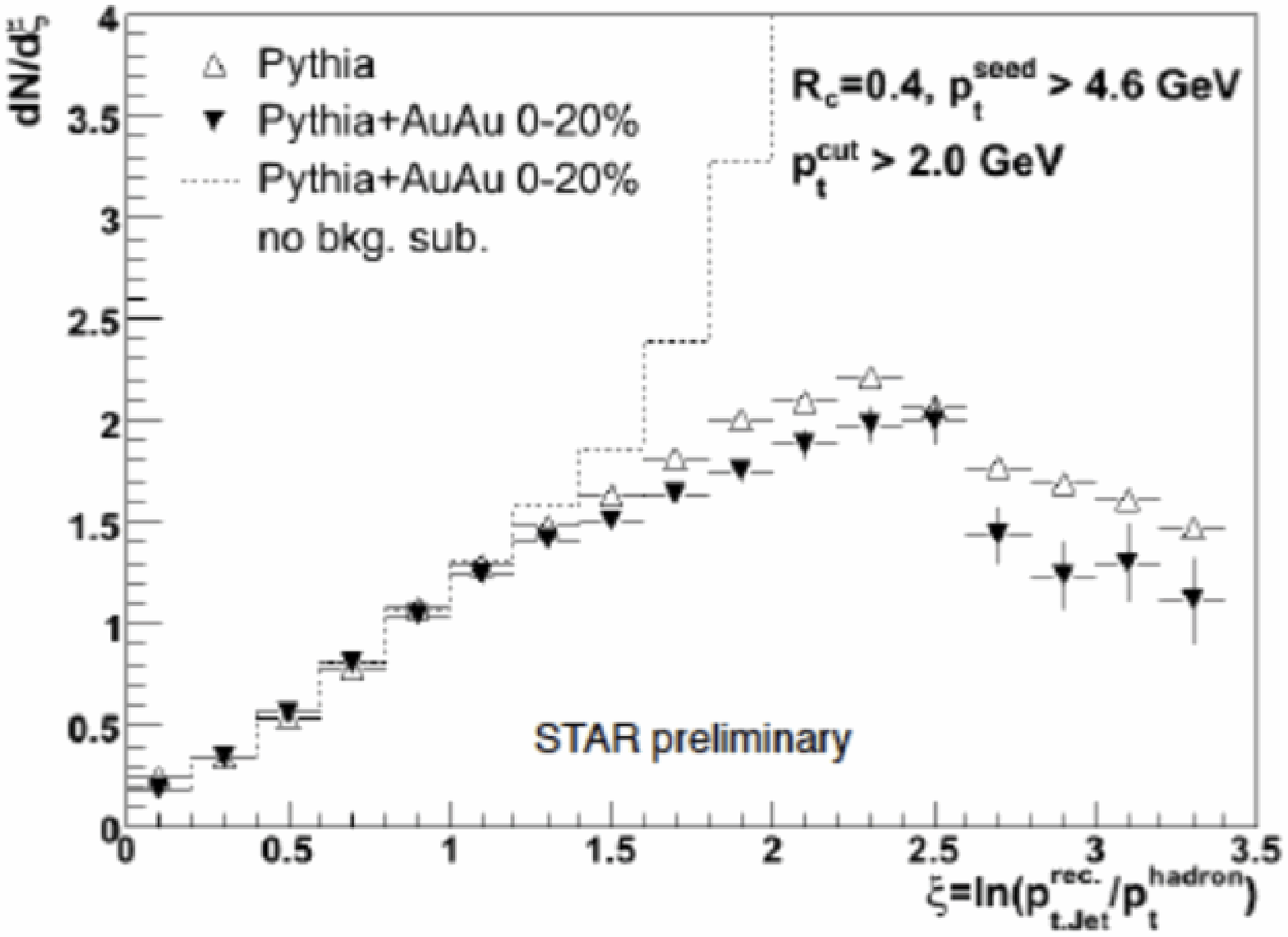}}
\resizebox{0.42\textwidth}{!}{  \includegraphics{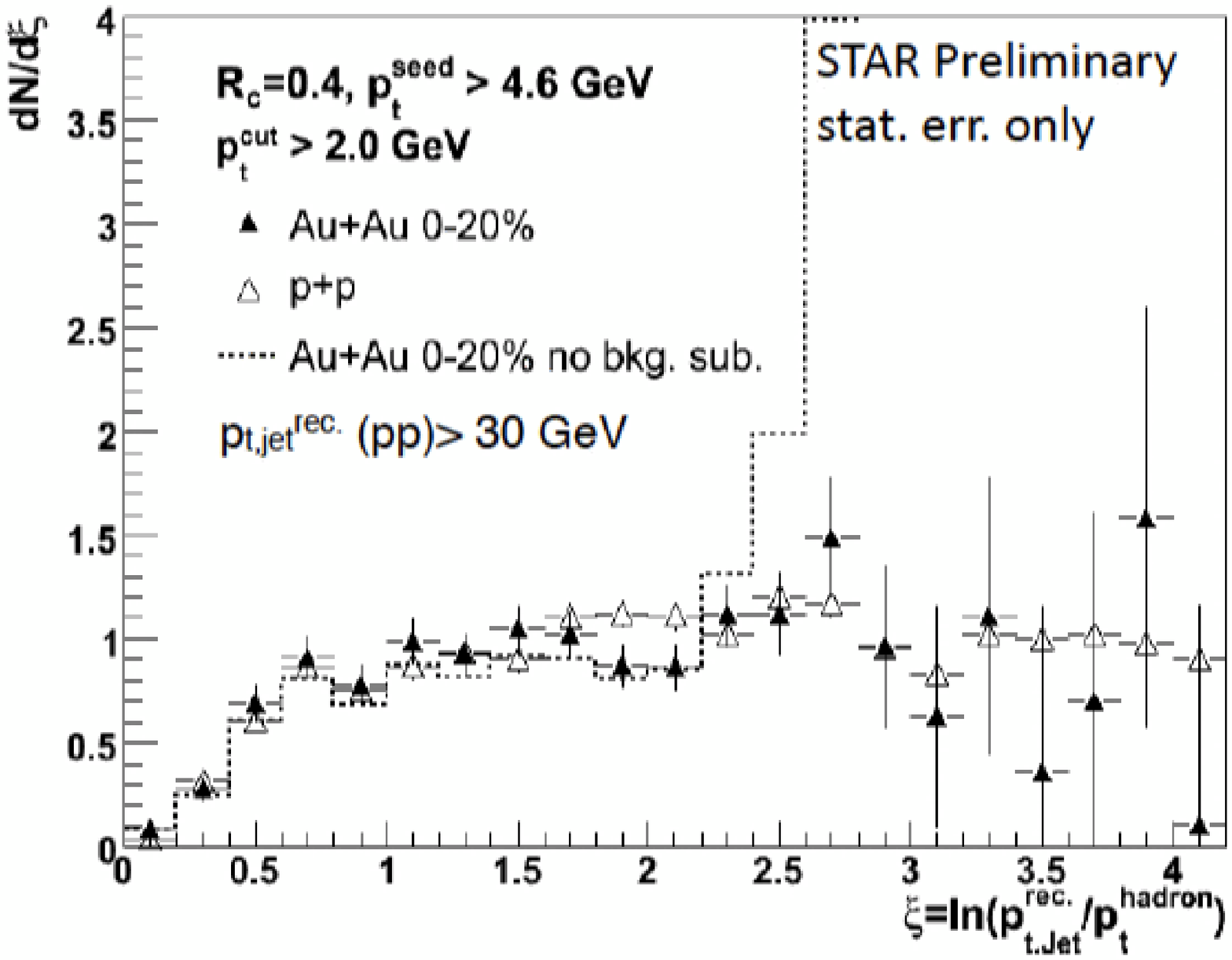}}

\vspace{-3mm}
\caption{Left plot: $\xi$ distributions of 30 GeV Pythia jets (open symbols), 30 GeV Pythia jets embedded in 0-20$\%$ MB Au+Au events with background subtraction (close symbols) and no background subtraction (dotted histogram). Right plot: uncorrected $\xi$ distributions measured in p+p HT (open symbols) and 0-20$\%$ Au+Au HT events at $\sqrt s_{NN}=200$ GeV for $p_{t,jet}^{rec}(pp)>30$ GeV and $p_t^{cut}=2$ GeV, with background subtraction (close symbols) and no background subtraction (dotted histogram). Only statistical errors.}
\label{fig:AuAuFF}
\end{figure}

\begin{figure}
\centering
\vspace{-5mm}
\resizebox{0.62\textwidth}{!}{  \includegraphics{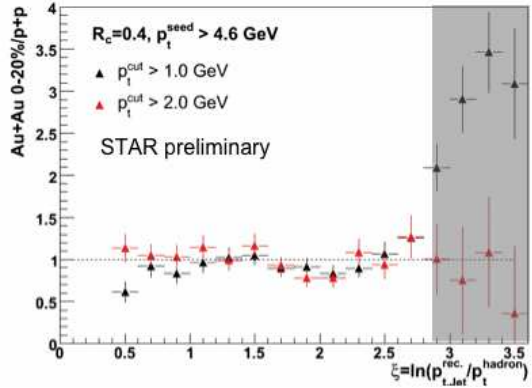}}
\vspace{-3mm}
\caption{Ratio of the $\xi$ distributions measured in 0-20$\%$ Au+Au HT events at $\sqrt s_{NN}=200$ GeV to p+p HT events at $\sqrt s=200$ GeV. The shaded area indicates the $\xi$ region where the uncertainties due to background subtraction are large.}
\label{fig:RatioFF}
\end{figure}

The jet energy is determined in R=0.4 and with a $p_t^{cut}$ to suppress the background and its fluctuations. In order to include also the soft jet fragments, the fragmentation functions are measured from charged hadrons in a radius of 0.7 around the jet axis without any minimum cut on transverse momentum of tracks. The measured fragmentation functions contain both jet and background components. The latter significantly contaminates the low momentum region of the fragmentation function and has to be subtracted.
The background component of the fragmentation function is estimated on an event-by-event basis from charged particle spectra in the out-of-cone area. %random cone areas outside the areas of the two jets with highest energy. 
As shown in Fig.~\ref{fig:AuAuFF} (left), the fragmentation functions of mono-energetic 30 GeV Pythia jets with and without embedding in Au+Au events are in agreement modulo small deviations due to the jet energy resolution: this suggests that the background subtraction procedure provides a reliable measurement especially of the low-$\xi$ region of the fragmentation functions assuming Pythia fragmentation.
The measured $\xi$ distributions in p+p and Au+Au HT events are shown in  Fig.~\ref{fig:AuAuFF} (right), for $p_{t,jet}^{rec}(pp)>30$ GeV. The corresponding jet momentum that selects similar jet populations in Au+Au as in p+p is $p_{t,jet}^{rec}(AuAu)>31$ GeV for $p_t^{cut}=2$ GeV and was estimated with Pythia simulations (as described in the previous session).
The background contribution from Au+Au jets with $p_{t,jet}^{rec}(AuAu)>31$ GeV (Fig.~\ref{fig:AuAuFF}, right) starts to be dominant for $\xi>2.5-3$. 
The ratio of the above fragmentation functions is shown in Fig.~\ref{fig:RatioFF} for two choices of $p_t^{cut}$. The ratio close to 1 may suggest that the fragmentation functions in Au+Au HT events are not  modified with respect to the reference p+p HT in the range  $0.5 \lsim \xi \lsim 2.5$. Further investigations are needed to understand the full implication of the result.

\section{Discussion of the results}
Results on jet spectra in Au+Au MB (\cite{sevil},\cite{sevilWW}) suggest a binary scaling of jet production with respect to p+p within large systematic errors.
With an unbiased jet population a softening of the measured fragmentation functions in the nuclear medium would be expected in order to explain the high-$p_T$ hadron suppression observed at RHIC~\cite{RaaSTAR,RaaSTAR2}.
On the contrary, no apparent modification of fragmentation functions is observed in Au+Au HT events. Therefore, use of the HT trigger together with a $p_t^{cut}$ seems to select a biased jet population.
Two effects may occur: first, the HT trigger in Au+Au may select jets that are similar to p+p because of ``surface'' effect and therefore are not modified by the medium. Second, the $p_t^{cut}=2$ GeV (chosen because similar jet populations are selected in p+p and Au+Au assuming Pythia fragmentation) could largely affect the measurement if jets are softened in the medium and as a result the reconstructed energy would be underestimated.
 
\section{Conclusions and outlook}
The measurements reported in these proceedings show that full jet reconstruction is feasible at RHIC, although especially challenging in the heavy-ion environment.
The jet fragmentation functions were measured in p+p with different jet-finding algorithms. The measurements provide a solid reference for Au+Au; they also suggest that out-of-cone radiation seems to be small at RHIC energies for cone radii R=0.4 and 0.7.
The fragmentation functions in Au+Au  for jet energies of 30 GeV and above were reported. Background fluctuations in Au+Au were reduced with a $p_t^{cut}$ on particles. No modification was observed in the $\xi$ distribution, indicating that a biased jet population was selected. 
A possible way to avoid the trigger bias is to study di-jets in the Au+Au HT events and measure fragmentation functions of recoil jets with respect to the trigger jets. In order to avoid biases due to the $p_t^{cut}$, the analysis can be done with no $p_t^{cut}$ on the recoil jet, but in this case background fluctuations will need to be unfolded to get fragmentation functions that can be directly compared to p+p.
In addition, given the complexity of such measurements, data driven correction scheme and systematic error estimates (fake jets, detector effects, etc) are needed to have more control on our understanding of the jet fragmentation functions in Au+Au. These analyses are currently in progress.

\end{document}